\documentclass[longbibliography,aps,prb,reprint,superscriptaddress]{revtex4-2}

\usepackage{amsmath}
\usepackage{amssymb}
\usepackage{graphicx}
\usepackage{xcolor}
\usepackage[hypertexnames=false]{hyperref}
\setcitestyle{numbers,square}

\newcommand{\revcolor}{black}
\newcommand{\rev}[1]{\textcolor{\revcolor}{#1}}

\begin{document}

\title{\rev{Data-Driven Thiele Equation Approach for State-Dependent Coefficients in the Nonlinear Dynamics of Vortex-based Nano-Oscillators}}

\author{Colin DUCARME}
\email{colin.ducarme@uclouvain.be}

\author{Simon DE WERGIFOSSE}

\author{Thomas G. COPP\'EE}

\author{Tristan \uppercase{da Câmara Santa Clara Gomes}}

\author{Flavio ABREU ARAUJO}
\email{flavio.abreuaraujo@uclouvain.be}

\affiliation{Institute of Condensed Matter and Nanosciences, Universit\'e catholique de Louvain, Louvain-la-Neuve, Belgium}

\begin{abstract}
Spin-torque vortex oscillators provide a model system for the nonlinear dynamics of a confined magnetic texture. Their motion is commonly described by the Thiele equation, but its standard form \rev{keeps the gyrovector and damping tensor fixed at their rigid-vortex values} and becomes inaccurate at large gyration amplitudes. We introduce a data-driven Thiele equation approach (DD-TEA) that extracts effective, position-dependent Thiele coefficients from a single current-ramp micromagnetic simulation by interpolating the magnetization in vortex-core-position space. The extracted maps reveal a weak increase of the gyrovector magnitude and a pronounced separation of the radial and azimuthal dissipation as the orbit expands, providing quantitative signatures of confinement- and motion-induced vortex deformation. Because the gyrotropic, dissipative, conservative, and spin-transfer contributions retain their usual Thiele structure, the resulting description remains physically interpretable rather than acting as a black-box surrogate. Incorporating these state-dependent coefficients into the equation reproduces the nonlinear stable-orbit dynamics and accurately predicts the response to time-varying currents, whereas a conventional constant-coefficient description predicts vortex expulsion. The framework therefore provides a systematic route for deriving effective collective-coordinate dynamics from full micromagnetic states with high computational efficiency.

DOI: \href{https://doi.org/10.1103/4dbr-s8mw}{10.1103/4dbr-s8mw}
\end{abstract}

\maketitle

\section{Introduction}

Magnetic vortices are nonuniform magnetization states in which the in-plane magnetization curls around a nanoscale core \cite{Shinjo2000}. Their low-frequency gyrotropic mode, governed primarily by the motion of this core \cite{Choe2004,Guslienko2006}, provides a canonical example of collective magnetization dynamics in a confined ferromagnet. When sustained by spin-transfer torque, this mode gives rise to spin-torque vortex oscillators (STVOs), which are promising nonlinear microwave sources \cite{Pribiag2007,Mistral2008,Dussaux2010} and building blocks for spintronic signal processing \cite{Torrejon2017,Romera2018}. Beyond these applications, their current tunability and rich nonlinear response \cite{Slavin2009,Hamadeh2014} make STVOs a controlled setting in which to investigate how the motion of a localized core couples to deformations of the surrounding magnetic texture.

Collective-coordinate approaches provide a natural language for this problem by projecting the full magnetization dynamics onto a small set of variables associated with the slow modes of the texture \cite{Thiele1973,Huber1982,Tretiakov2008,Metlov2013}. \rev{Formally, such reductions follow from the Lagrangian formulation of magnetization dynamics introduced by D\"oring \cite{Doring1948} and Gilbert \cite{Gilbert2004}, in which the Landau--Lifshitz--Gilbert (LLG) equation derives from a Lagrangian supplemented by a Rayleigh dissipation function, and in which any parametrization of the texture can serve as a set of generalized coordinates; a Hamiltonian formulation also exists \cite{Papanicolaou1991}, although dissipation is difficult to accommodate within it. The Thiele equation was originally derived for the particular choice of a rigidly translating texture, its displacement being the only coordinate, but the rigid hypothesis is not required by the equation: for any deformation of the form $\mathbf m(\mathbf r;\mathbf X_\mathrm{c})$ (with $\mathbf{X}_\mathrm{c}$ the vortex-core position), the Lagrangian prescription yields the same force-balance equation, with the gyrotropic, dissipative, and conservative coefficients given by the standard integrals over the tangent fields $\partial\mathbf m/\partial\mathbf X_\mathrm{c}$ \cite{Tretiakov2008}. Rigid translation is the special case in which these tangent fields reduce to the spatial derivatives of a fixed texture.} For a magnetic vortex, the usual coordinate is the vortex-core position $\mathbf{X}_\mathrm{c}$, and the corresponding Thiele equation expresses a balance among gyrotropic, dissipative, conservative, and current-induced forces \cite{Guslienko2002,Antos2008}. Its appeal is not only that it reduces the dimensionality of the \rev{LLG} dynamics, but also that each term retains a direct physical interpretation. The gyrotropic constant characterizes the transverse response of the texture, the damping tensor describes the dissipative metric associated with collective motion, the restoring force connects the core trajectory to the magnetic energy landscape, and the external forces connects it to applied torques, such as \rev{spin-transfer}, spin-orbit, etc.

\rev{In practice, the gyrovector and damping tensor are almost always assigned the constant values that follow from a rigidly translating vortex \cite{Huber1982,Guslienko2006_APL}. Applying the same rigid ansatz to the potential energy gives a poor estimate of the gyrotropic frequency, and since the work of Guslienko \textit{et al.} \cite{Guslienko2002,Guslienko2006} the mainstream description has therefore been a hybrid one: rigid-vortex kinetic and dissipative terms combined with a deformable two-vortex ansatz \cite{Metlov2002} for the magnetostatic energy or, more recently, with a position-dependent restoring force obtained from micromagnetic simulations \cite{deWergifosse2023}. Metlov \cite{Metlov2013} identified this reliance on two different models for the kinetic and potential terms as a limitation and proposed to remove it analytically, by inserting the same non-rigid two-vortex ansatz into both terms of the micromagnetic Lagrangian and by taking the parameters of the ansatz, rather than the displacement of the core, as collective coordinates. Any prescribed ansatz nevertheless fixes in advance how the texture may deform, which becomes restrictive in the nonlinear regime.} As the core approaches the disk edge, confinement modifies the surrounding magnetization, while the core velocity produces additional dynamical deformations \cite{Vansteenkiste2009,Dussaux2012,Yoo2021}. Nonlinear gyrotropic motion, frequency shifts, and departures from rigid-vortex predictions have consequently been observed in micromagnetic and experimental studies \cite{Lee2007,Khvalkovskiy2009,Guslienko2010Nonlinear,Drews2012}. Dynamic corrections to the gyrovector and damping tensor have been proposed to account for such motion \cite{Shim2011}, whereas extended collective-coordinate descriptions associate internal deformations with inertial or mass-like effects \cite{Volkel1994,Wysin1996,Yoo2021}. These results show that a constant-coefficient Thiele equation can miss not only quantitative details but also the manner in which different parts of the force balance evolve with the state of a deformable texture.

Full micromagnetic simulations (MMS) contain these effects without imposing a rigid ansatz, but they do not by themselves provide a compact coefficient-level description of the resulting dynamics. Conversely, machine learning-based models can be predictive while sacrificing the interpretation supplied by a collective-coordinate equation \cite{Chen2022}. The central question addressed here is therefore how effective collective-coordinate descriptions can be systematically extracted from full micromagnetic dynamics while retaining physical interpretability.

\begin{figure}
\centering
\includegraphics[width=0.45\textwidth]{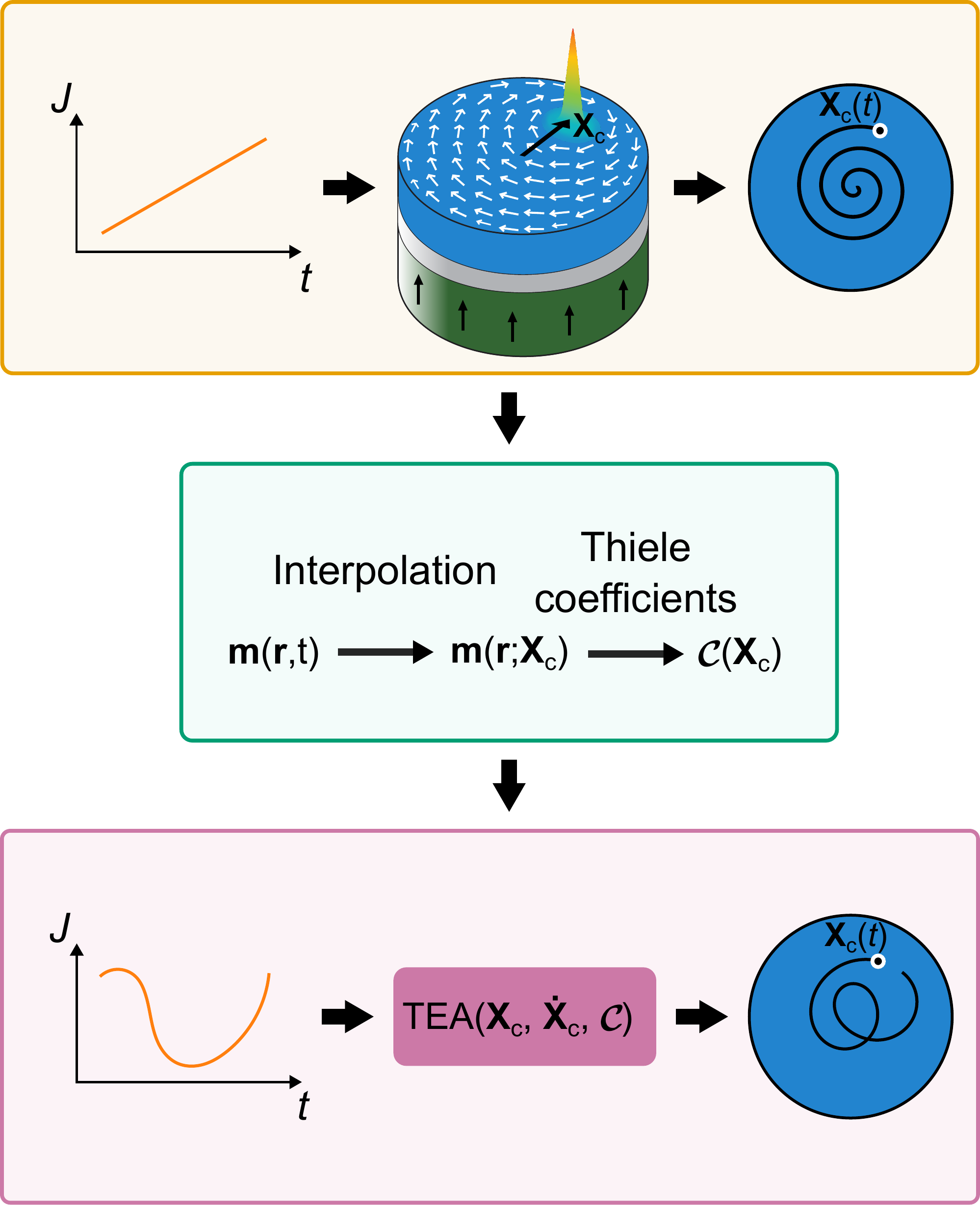}
\caption{State-dependent Thiele-equation workflow. (Orange) A single current-ramp micromagnetic simulation generates vortex states spanning a broad range of vortex-core positions $\mathbf{X}_\mathrm{c}$. (Green) The magnetization field is reparameterized in core-position space, defining a sampled manifold from which the effective Thiele coefficients $\boldsymbol{\mathcal{C}}$ are extracted. (Pink) The resulting state-dependent force balance is used to predict vortex dynamics under independent current protocols and is compared with full micromagnetic simulations.}
\label{fig:pipeline}
\end{figure}

We introduce a data-driven Thiele equation approach (DD-TEA) that extracts position-dependent effective Thiele coefficients directly from micromagnetic snapshots. The magnetization is interpolated in vortex-core-position space, so that the gyrotropic, dissipative, conservative, Oersted-field, and spin-transfer contributions can be evaluated on the manifold of states visited by the vortex. The extracted coefficients are not regarded as material constants: they are effective quantities associated with the chosen collective coordinate and encode the deformation of the full magnetization texture as the vortex state changes. \rev{The approach thus shares the diagnosis of Ref.~\cite{Metlov2013}, that kinetic and potential terms should derive from one and the same description of the texture, but it reaches this consistency without an ansatz: both kinds of terms are evaluated on the micromagnetic states themselves, at the price of retaining, for now, the core position as the only collective coordinate.} The resulting equation retains the structure of the Thiele force balance and is therefore distinct from a black-box surrogate, while its reduced dimensionality also makes subsequent trajectory calculations inexpensive. The overall micromagnetic-to-Thiele workflow is summarized in Fig.~\ref{fig:pipeline}.

We show that, over the operating regime in which the vortex state exists and the core polarity is preserved, a single current-ramp simulation provides sufficient information to construct an accurate reduced description. The extracted maps reveal a \rev{core-position-dependent} gyrotropic response and an increasing separation between radial and azimuthal dissipation, directly exposing the breakdown of a rigid, constant-coefficient picture. Incorporating these variations reproduces the nonlinear stable-orbit dynamics and predicts transient responses under time-dependent drives beyond the training trajectory, whereas a conventional constant-coefficient model predicts vortex expulsion. Although demonstrated for STVOs, the framework provides a general route for constructing interpretable collective-coordinate equations for deformable magnetic textures directly from micromagnetic states.

\section{State-dependent collective-coordinate framework}

Micromagnetic simulations were performed using mumax$^+$ \cite{Moreels2026} for a Permalloy disk of radius $R=100$ nm and thickness $h=10$ nm. The reduced description is constructed by sampling the magnetization over a broad range of vortex-core positions. All material and simulation parameters are provided in the Supplemental Materials.

For each saved snapshot, the vortex core is located and its position is written as
\begin{equation}
\mathbf{X}_\mathrm{c}=X_\mathrm{c}\hat{\mathbf{x}}+Y_\mathrm{c}\hat{\mathbf{y}}
=sR\,\hat{\mathbf e}_r(\varphi),
\end{equation}
where $s=|\mathbf X_\mathrm{c}|/R$ is the normalized orbit radius, and $\varphi$ is the core azimuthal angle. The aim is not to impose an analytical ansatz for the magnetization texture, but to identify the effective force balance associated with the micromagnetic states visited by the system.

Each micromagnetic snapshot provides the magnetization field
\begin{equation}
\mathbf m=\mathbf m(\mathbf r;\mathbf X_\mathrm{c}),
\end{equation}
which is regarded as a function of both the spatial coordinate $\mathbf r$ and the collective coordinate $\mathbf X_\mathrm{c}$. The set of snapshots generated during the current ramp therefore forms a discrete sampling of a magnetization-state manifold over the $(X_\mathrm{c},Y_\mathrm{c})$ plane. For every spatial cell, the magnetization is interpolated between neighboring sampled core positions, yielding a continuous representation of $\mathbf m(\mathbf r;\mathbf X_\mathrm{c})$ in core-position space. The derivatives $\partial\mathbf m/\partial X_\mathrm{c}$ and $\partial\mathbf m/\partial Y_\mathrm{c}$ are then tangent fields to this manifold: they describe how the complete magnetic texture changes under an infinitesimal displacement of the selected collective coordinate, rather than assuming that such a displacement is a rigid translation. \rev{By construction these integrals are taken over derivatives with respect to the collective coordinates rather than over a rigid translation of a fixed texture; they therefore realize the generalized, non-rigid form of the Thiele equation \cite{Tretiakov2008}, which remains valid for any deformation of the texture that is a single-valued function of the core position along the sampled manifold.}

The effective Thiele coefficients are obtained by evaluating the standard Thiele integrands locally on each cell and summing their contributions over the magnetic volume. In general, each coefficient $\mathcal C_\alpha$ can be computed as
\begin{equation}
\mathcal C_\alpha(\mathbf X_\mathrm{c})
=
\int_V
f_\alpha\!\left(
\mathbf m,
\frac{\partial\mathbf m}{\partial X_\mathrm{c}},
\frac{\partial\mathbf m}{\partial Y_\mathrm{c}}
\right)\,
\mathrm dV,
\end{equation}
where the function $f_\alpha$ depends on the coefficient under consideration, as specified in the Supplemental Materials. This procedure is applied to the gyrotropic coefficient $G$, the damping-tensor components $D_{ij}$, the Oersted-field contribution $\kappa_{\mathrm{Oe}}$, and the spin-transfer efficiency $\kappa_{\perp}^{\mathrm{ST}}$. The gyrotropic and dissipative coefficients probe, respectively, the antisymmetric and symmetric couplings between the tangent fields, while the remaining terms quantify the forces produced by the energy landscape and current-induced torques. Their state dependence therefore has a direct interpretation in terms of how the full texture changes along the collective-coordinate manifold.

The extracted quantities are assembled into an effective Thiele equation whose coefficients depend on the vortex-core position. Following the Thiele formalism \cite{Thiele1973,Huber1982,AbreuAraujo2022}, the reduced dynamics is written as
\begin{equation}
\begin{aligned}
-P|G|\,(\hat{\mathbf z}\times\dot{\mathbf X}_\mathrm{c})
&+\mathbf D\cdot\dot{\mathbf X}_\mathrm{c}
\\
=
\Bigl[k_0+CJ\,\kappa_{\mathrm{Oe}}\Bigr]\mathbf X_\mathrm{c}
&+\kappa_{\perp}^{\mathrm{ST}}\,
J\,(\hat{\mathbf z}\times\mathbf X_\mathrm{c}),
\end{aligned}
\label{eq:thiele}
\end{equation}
where $P$ and $C$ are respectively the vortex polarity and chirality, and $J$ is the current density. The equation has the same force-balance structure as a conventional Thiele model; the difference is that its coefficients are evaluated on the micromagnetically sampled state manifold rather than \rev{derived from a prescribed ansatz for the texture, whether rigid for all terms or, as in the hybrid descriptions discussed in the Introduction, rigid for the kinetic and dissipative terms and deformable for the potential energy}.

For a circular nanodot, cylindrical symmetry implies that the damping tensor is diagonal in the polar basis $(\hat{\mathbf e}_r,\hat{\mathbf e}_\varphi)$,
\begin{equation}
\mathbf D
=
D_{rr}\,\hat{\mathbf e}_r\otimes\hat{\mathbf e}_r
+
D_{\varphi\varphi}\,\hat{\mathbf e}_\varphi\otimes\hat{\mathbf e}_\varphi,
\label{eq:D}
\end{equation}
so that the dissipative response to radial relaxation and azimuthal phase motion can be resolved separately. Although the microscopic Gilbert damping remains scalar in the simulations, these two collective motions need not produce the same spatial rearrangement of the magnetization and can therefore acquire different effective damping coefficients.

Repeating the extraction for all saved frames along the ramp yields discrete values of $G$, $D_{rr}$, $D_{\varphi\varphi}$, $k_0$, $\kappa_{\mathrm{Oe}}$, and $\kappa_{\perp}^{\mathrm{ST}}$ across the explored orbit range. For an ideal circular nanodot, smooth scalar coefficients are rotational invariants and can be represented as analytic functions of $|\mathbf X_\mathrm{c}|^2$. We therefore fit them using even powers of $s$ up to order eight. The fits provide a compact representation of the state dependence and are inserted into Eq.~\eqref{eq:thiele} to reconstruct vortex trajectories. In contrast to a purely trajectory-based surrogate, the interpolation is used only to recover the magnetization manifold from which physically identified Thiele terms are evaluated. The resulting coefficients should consequently be viewed as effective parameters of the chosen collective-coordinate description, not as intrinsic material constants.

\section{State-dependent gyrotropic and dissipative dynamics}\label{sec:extraction}

\begin{figure}
\centering
\includegraphics[width=0.48\textwidth]{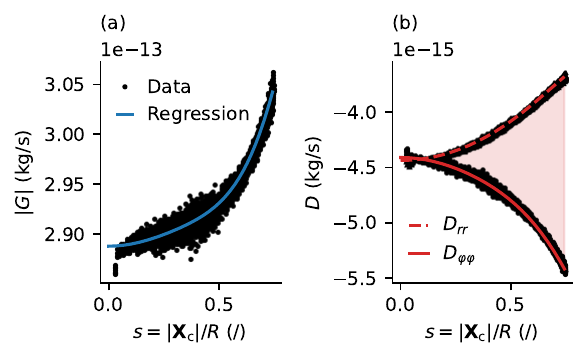}
\caption{Extracted effective Thiele coefficients as functions of the normalized orbit radius $s$. (a) Gyrotropic coefficient $|G(s)|$. (b) Radial and azimuthal damping coefficients, $D_{rr}(s)$ and $D_{\varphi\varphi}(s)$. Symbols denote values extracted from the micromagnetic snapshots, while lines show the polynomial regressions used in the reduced description. The radius dependence of $G$ and the increasing separation between $D_{rr}$ and $D_{\varphi\varphi}$ quantify the departure of the collective dynamics from a rigid, constant-coefficient vortex model.}
\label{fig:coefficients}
\end{figure}

Figure~\ref{fig:coefficients} summarizes the effective coefficients extracted from the current-ramp simulation. The polynomial regressions are restricted to the range $s\leq s_\mathrm{max}-l_\mathrm{ex}/R$, with $s_\mathrm{max}$ the last orbit radius visited and $l_\mathrm{ex}$ the exchange length. This cutoff corresponds to the onset of a strongly nonlinear regime in which the vortex approaches the disk edge and the magnetization texture undergoes increasingly rapid deformations \cite{Khvalkovskiy2009}. Beyond this point, the extracted coefficients exhibit a marked change in behavior as the system approaches the dynamical core-reversal threshold, which occurs near $s_\mathrm{max}\approx0.8$ for the considered nanodot \cite{AbreuAraujo2022}. Restricting the fit to a length $l_\mathrm{ex}$ before reversal therefore defines a state manifold composed of stable vortices while retaining the large-amplitude configurations needed to probe nonlinear collective dynamics.

Within the fitted range, the even-order polynomial regression provides an excellent description of the micromagnetic data, with $R^2=0.975$ for $G$, $0.998$ for $D_{rr}$, and $0.995$ for $D_{\varphi\varphi}$. The extracted coefficients are therefore represented by smooth functions that remain well constrained by the simulated states over the explored range of vortex-core positions.

The gyrotropic coefficient is commonly connected to the topological charge of the vortex and is consequently treated as constant in simplified Thiele descriptions \cite{Tretiakov2008}. This constancy follows directly for a rigid translation, for which derivatives with respect to the core position reduce to spatial derivatives of a fixed texture. In the present extraction, however, $\partial\mathbf m/\partial\mathbf X_\mathrm{c}$ also contains the change in vortex shape produced by confinement and motion. The resulting $G(s)$ is therefore the effective gyrotropic coupling of the selected core coordinate to the complete, deformable texture. Its magnitude increases by approximately $8\%$ over the fitted range. This variation does not imply a change of core polarity or topological sector; instead, it quantifies the progressive departure of the sampled state manifold from rigid translation, consistent with dynamic corrections to the gyrovector reported for nonlinear vortex motion \cite{Shim2011}.

The damping tensor exhibits the strongest position dependence. Both components coincide at small orbit radius, consistent with a nearly rigid and rotationally symmetric response near the disk center. As the orbit expands, however, the two coefficients progressively separate. The magnitude of $D_{rr}$ decreases by approximately $20\%$, whereas $D_{\varphi\varphi}$ increases by approximately $25\%$. Since $D_{ij}$ is determined by the spatial overlap of the tangent fields $\partial_i\mathbf m$ and $\partial_j\mathbf m$, this separation measures an anisotropy of the collective-coordinate manifold: a unit radial displacement and a unit azimuthal displacement no longer induce equivalent rearrangements of the magnetization. The effect is therefore an emergent anisotropy of the collective dissipation, despite the scalar Gilbert damping used in the micromagnetic model. A constant isotropic damping coefficient cannot represent this distinct evolution of radial relaxation and phase dynamics.

The remaining coefficients, namely $k_0(s)$, $\kappa_{\mathrm{Oe}}(s)$, and $\kappa_{\perp}^{\mathrm{ST}}(s)$, are also extracted and fitted. The stiffness terms have already been discussed extensively in previous work \cite{deWergifosse2023}, and their corresponding curves are provided in the Supplemental Materials. In contrast to the gyrotropic and dissipative terms, $\kappa_{\perp}^{\mathrm{ST}}$ remains nearly constant over the explored orbit range. This comparison is physically informative: the direct spin-transfer efficiency remains close to its rigid-vortex value, while the effective transverse and dissipative responses are substantially modified as the texture deforms. All coefficients are nevertheless retained in the reduced equation and contribute to the predicted dynamics.

The extracted quantities should not be interpreted as material constants. During the current-driven oscillations used for the extraction, the dominant source of deformation is correlated with the orbit radius: as the core approaches the disk edge, the surrounding magnetization is progressively distorted by confinement \cite{Dussaux2012,Vansteenkiste2009}. A second contribution arises from the vortex velocity \cite{Yoo2021}. For steady-state gyrotropic motion, the orbit radius determines the oscillation frequency and therefore the core velocity, so larger orbits are naturally associated with stronger dynamical distortions \cite{AbreuAraujo2022}. The one-dimensional maps $\mathcal C(s)$ consequently provide an effective closure in which internal deformation degrees of freedom are slaved to the orbit radius along the sampled dynamics.

Extended collective-coordinate descriptions instead introduce deformation amplitudes explicitly, producing inertial or mass-like terms and velocity-dependent effective masses \cite{Volkel1994,Wysin1996,Yoo2021}. The present first-order, state-dependent description is complementary to these approaches. It absorbs the quasi-steady influence of velocity and deformation into the coefficient maps, but it does not assert that inertia is absent. If the same core position can be reached with different velocities or different internal states, a single-valued map in $s$ is no longer sufficient. This distinction becomes important for the transient validation discussed below and defines a direct route toward higher-dimensional collective-coordinate descriptions.

\rev{The maps in Fig.~\ref{fig:coefficients} are specific to the simulated geometry and material. Changing the dot dimensions or the exchange length, or adding interactions such as magnetocrystalline anisotropy or the Dzyaloshinskii--Moriya interaction, changes the sampled state manifold and hence the extracted coefficients, but not the extraction itself, which only requires micromagnetic snapshots labeled by the core position. What such changes can affect is the dimensionality of the maps. The reduction to functions of $s$ alone relies on the cylindrical symmetry of the dot; a geometry or an interaction that breaks this symmetry would require coefficients depending on both $s$ and $\varphi$, evaluated over the $(X_\mathrm{c},Y_\mathrm{c})$ plane that the ramp trajectory samples. The procedure is, however, restricted to a single magnetic phase. Parameter combinations for which the vortex state becomes unstable, or drives that lead to core reversal, spin-wave emission, or a transition to another magnetic state, are not captured, and the cutoff $s\leq s_\mathrm{max}-l_\mathrm{ex}/R$ introduced above enforces this restriction by excluding the approach to the reversal threshold from the fit.}

\section{Dynamics beyond the sampled trajectory}

The extracted coefficient functions are intended not only to reproduce the current-ramp data, but also to test whether the sampled state manifold defines a transferable collective-coordinate equation. We therefore compare the reduced description with full micromagnetic simulations for a composite time-dependent current protocol that was not used during extraction. The validation simulation is performed for the opposite vortex polarity and current sign relative to the training run. Throughout the validation, all Thiele coefficients remain fixed to the functions obtained from the single current-ramp simulation, with no additional fitting.

\begin{figure}
\centering
\includegraphics[width=0.48\textwidth]{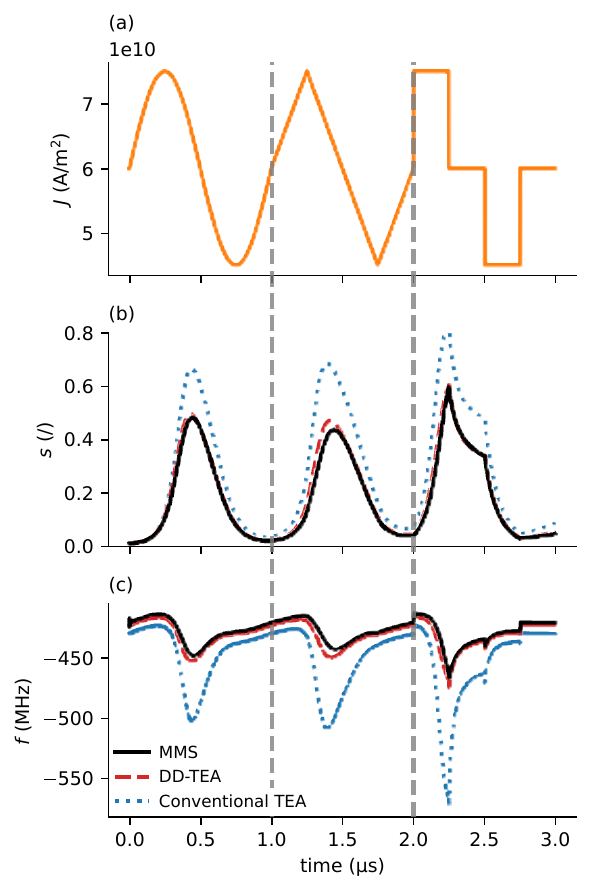}
\caption{Validation of the state-dependent Thiele description against independent micromagnetic simulations. (a) Applied current density $J(t)$. (b) Normalized orbit radius $s(t)$. (c) Instantaneous frequency $f(t)=\dot{\varphi}(t)/(2\pi)$. Full micromagnetic simulations (MMS) are compared with DD-TEA and with \rev{the conventional Thiele equation of the Supplemental Materials, which combines constant rigid-vortex gyrotropic, dissipative, and current-induced coefficients with a position-dependent restoring force}.}
\label{fig:validation}
\end{figure}

The applied current density is $J(t)=6.0\times10^{10}+1.5\times10^{10}\,u(t)$ A/m$^2$, where the waveform $u(t)$ combines sinusoidal, triangular, and step-like segments [Fig.~\ref{fig:validation}(a)]. This composite excitation probes the collective dynamics under smooth, piecewise-linear, and discontinuous changes of the drive within a single simulation.

Figure~\ref{fig:validation} compares the normalized orbit radius $s(t)$ and the instantaneous frequency $f(t)=\dot{\varphi}(t)/(2\pi)$, with $\varphi$ unwrapped before differentiation, obtained from MMS, DD-TEA, and the conventional constant-coefficient Thiele equation detailed in the Supplemental Materials. The conventional description predicts vortex-core expulsion, \textit{i.e.}, $s\geq0.8$, at $2.25~\mu$s. This failure is not caused by an inaccurate nonlinear stiffness because the reference model already includes a position-dependent restoring term. \rev{The reference model is thus representative of the hybrid descriptions discussed in the Introduction, in which only the potential energy departs from the rigid-vortex picture.} Rather, keeping the gyrovector and damping tensor fixed prevents the transverse and dissipative parts of the force balance from evolving with the deformed vortex state. The resulting error accumulates in both the radial relaxation and the phase dynamics until the model reaches a qualitatively incorrect expulsion trajectory.

In contrast, DD-TEA reproduces the overall transient response throughout the entire protocol. Quantitatively, it yields a mean-normalized RMSE of $9\%$ for $s(t)$ and $1\%$ for $f(t)$, compared with $51\%$ and $7\%$, respectively, for the conventional description. The agreement is retained despite reversing the vortex polarity and current sign relative to the training simulation. This transfer is consistent with the interpretation of the fitted functions as state-dependent properties of the vortex manifold within the symmetry of the circular dot, rather than as a direct interpolation of the training time series.

The remaining discrepancies are concentrated during the fastest transients. In these intervals, the instantaneous vortex velocity can depart significantly from the quasi-steady relation between orbit radius and velocity sampled during the ramp. Because the present coefficients depend only on $s$, velocity-induced deformations are absorbed into a one-dimensional parametrization and cannot vary independently. The residual mismatch therefore identifies the physical information omitted by the chosen coordinate set: rapid motion can excite deformation or inertial responses that are not uniquely specified by the instantaneous core position. A natural extension is to construct coefficient maps in $(\mathbf X_\mathrm{c},\dot{\mathbf X}_\mathrm{c})$ space or to introduce explicit deformation coordinates, allowing position- and velocity-dependent contributions to be separated. We are currently investigating such higher-dimensional parametrizations, and their importance should be taken into account in future studies aiming at predictive reduced-order models under more general excitation conditions.

The reduced dimensionality also has an important computational consequence. The reference MMS validation simulation required approximately $60$ hours on an NVIDIA GeForce RTX 3090, while the single current-ramp MMS used for extraction required approximately $20$ hours. Once this dataset has been generated, coefficient extraction and each subsequent DD-TEA prediction require only a few minutes on a single CPU core. The speedup is therefore a consequence of replacing repeated field-level evolution by an interpretable equation on the extracted state manifold.

\section{Discussion}

The extracted equation is an effective description tied to the selected collective coordinates. In the present case, the vortex-core position is sufficient when the remaining degrees of freedom are uniquely slaved to $\mathbf X_\mathrm{c}$ along the sampled motion. If two micromagnetic states have the same core position but different deformations, the mapping $\mathbf m(\mathbf r;\mathbf X_\mathrm{c})$ becomes multivalued and a position-only equation cannot distinguish them. The choice of coordinates is therefore part of the physical model rather than a purely numerical detail. This dependence also provides a diagnostic: systematic residuals indicate that an internal mode relevant to the force balance has not been resolved.

Velocity-dependent effects are the most immediate example. During the current ramp, orbit radius, frequency, and velocity remain strongly correlated, allowing their combined influence to be represented by $\mathcal C(s)$. External magnetic fields or sufficiently rapid current transients can modify the velocity independently of the radius, breaking this one-to-one relation. Higher-dimensional maps depending on $\mathbf X_\mathrm{c}$ and $\dot{\mathbf X}_\mathrm{c}$, or explicit collective coordinates describing core and background deformation, would then separate these contributions. Such extensions would connect the present state-dependent first-order equation to inertial descriptions in which internal deformation produces mass-like dynamics \cite{Volkel1994,Wysin1996,Yoo2021}. Rather than replacing those models, the extraction procedure offers a systematic way to determine when additional coordinates are required and to evaluate their effective couplings from micromagnetic states.

\rev{A related limitation concerns systems of strongly interacting vortices. Their mutual interaction modifies not only the potential energy but also the kinetic and dissipative terms, and the collective coordinates can no longer be the displacement of a single rigid texture. Within the present framework, the natural extension is to enlarge the coordinate set to the positions of all cores, $\mathbf X=(\mathbf X_1,\mathbf X_2,\dots)$, and to extract the tangent fields $\partial\mathbf m/\partial\mathbf X_k$ from snapshots of the coupled system, so that mutual deformation enters the gyrotropic and dissipative couplings in the same way as confinement does here. A simpler approximation would treat each vortex as driven by the time-dependent mean field produced by the others. Both routes require coefficients that respond to a field or a drive varying independently of the orbit radius, which is precisely the situation in which the one-dimensional maps $\mathcal C(s)$ are insufficient. Multi-vortex dynamics therefore lies outside the scope of the present extraction, and its treatment is tied to the velocity- and field-dependent extensions discussed above.}

The same perspective can be applied to other magnetic textures provided that a suitable set of collective coordinates can be identified. Domain walls, skyrmions, and hopfions are already described by collective-coordinate or Thiele-type equations in appropriate regimes \cite{Thiaville2006,Tretiakov2008,Makhfudz2012,Schuette2014,Wang2019}. For these systems, coordinates associated with position may need to be supplemented by width, size, helicity, or deformation modes, depending on the excitation and geometry. Constructing coefficient maps in these enlarged spaces would allow the role of deformability and inertia to be examined without abandoning the interpretability of the underlying force balance.

\rev{These extensions can be placed in the broader context of the Lagrangian and Hamiltonian formulations of magnetization dynamics \cite{Doring1948,Gilbert2004,Papanicolaou1991}. In the Lagrangian formulation, any set of generalized coordinates yields an equation of motion whose gyrotropic, dissipative, and conservative terms are integrals over the texture and over its derivatives with respect to those coordinates. Ref.~\cite{Metlov2013} exploits this freedom with an analytical ansatz whose parameters serve as coordinates and obtains equations that reproduce the LLG dynamics as far as the choice of collective variables allows. The extraction presented here evaluates the same kind of integrals, restricted to the core position, on micromagnetic states instead of an ansatz. Because these integrals are taken over the tangent fields $\partial\mathbf m/\partial\mathbf X_\mathrm{c}$, not over spatial derivatives of a rigid texture, they correspond to the generalized, non-rigid form of the Thiele equation \cite{Tretiakov2008} and apply unchanged to a deforming vortex. Along the near-steady current ramp, the core velocity is a single-valued function of the core position, so the deformations induced by confinement and by motion are single-valued functions of $\mathbf X_\mathrm{c}$ on the sampled manifold; the state-dependent coefficients consequently encode them exactly. It is this property, rather than any rigidity assumption, that makes the reduced equation accurate. It does not aim at an equivalent reformulation of the LLG equation: its purpose is a reduced equation that is accurate enough for predictive use at the cost of a Thiele-equation integration, without the need for an approximate analytical texture. Enlarging the coordinate set within this data-driven setting, as discussed above, is the counterpart of the flexible choice of coordinates in the analytical approaches.}

Finally, the present maps are constrained to the stable-vortex region sampled by the training ramp and are not intended to extrapolate through core reversal or into a different magnetic state. \rev{Changes of geometry or material parameters modify these manifolds and their boundaries, and therefore the extracted maps, but not the extraction procedure itself (Sec.~\ref{sec:extraction}).} This restriction defines an opportunity for piecewise collective-coordinate descriptions in which separate state manifolds are extracted for distinct dynamical regimes. Likewise, incorporating thermal fluctuations would require either ensemble-averaged coefficients or a stochastic collective-coordinate equation. These extensions preserve the central principle of the approach: the reduced equation and its domain of validity should be inferred from the structure of the micromagnetic state manifold.

\section{Conclusion}

In conclusion, this work provides a physically interpretable route for extracting effective dynamical equations directly from micromagnetic simulations and offers new insight into nonlinear vortex dynamics. By evaluating the Thiele terms on a micromagnetically sampled vortex-state manifold, the method retains the gyrotropic, dissipative, conservative, and spin-transfer structure of the collective-coordinate equation without imposing a rigid analytical texture.

The extracted coefficients show that the nonlinear dynamics cannot be reduced to a position-dependent restoring force alone. The gyrovector magnitude increases with orbit radius, while the radial and azimuthal damping coefficients evolve in opposite directions, revealing an emergent anisotropy of the collective dissipation. These trends quantify how confinement- and motion-induced deformation modifies the effective force balance even though the vortex polarity is preserved and the microscopic Gilbert damping remains scalar.

Including this state dependence reproduces stable nonlinear orbits and predicts the response to an independent time-varying current protocol, whereas a conventional constant-coefficient description predicts vortex expulsion. The remaining transient discrepancies identify velocity-dependent deformation as a natural target for higher-dimensional coefficient maps or additional collective coordinates. The resulting reduction in computational cost is substantial, but it follows from the more fundamental result: full micromagnetic dynamics can be converted into an interpretable, state-dependent collective-coordinate equation for a deformable magnetic texture.

\section*{Acknowledgements}
Computational resources have been provided by the super-computing facilities of the Université catholique de Louvain (CISM/UCL) and the Consortium des Équipements de Calcul Intensif en Fédération Wallonie Bruxelles (CÉCI) funded by the Fond de la Recherche Scientifique de Belgique (F.R.S.-FNRS) under convention 2.5020.11 and by the Walloon Region. F.A.A. acknowledges support as a Research Associate of the F.R.S.-FNRS, and C.D. acknowledges support as a FRIA grantee of the F.R.S.-FNRS and from F.A.A. T.G.C. acknowledges support from Sirris through the EU CHIPS JU Joint Undertaking under Grant Agreement No. 101194414 (project MOSAIC). \\
The study was designed by F.A.A. and C.D., who created the analytical model. C.D. performed the simulations and wrote the core of the manuscript. All authors (C.D., S.d.W., T.G.C., T.d.C.S.C.G., and F.A.A.) contributed to the analysis and writing. \\
The authors have no conflicts to disclose. The funders had no role in study design, data collection, data analysis, interpretation, or manuscript writing.

\section*{Data Availability}
The dataset and visualization scripts supporting the findings of this study are publicly available on Zenodo at \href{https://doi.org/10.5281/zenodo.21398269}{10.5281/zenodo.21398269}. Additional simulation and analysis files are available from the corresponding author upon reasonable request.

\nocite{Moreels2026, Dussaux2012, Guslienko2014, Alfeld1984, Thiele1973, AbreuAraujo2022, Guslienko2011, deWergifosse2023, Guslienko2002, Guslienko2006_APL, Khvalkovskiy2009} 

%


\begin{thebibliography}{42}%
\makeatletter
\providecommand \@ifxundefined [1]{%
 \@ifx{#1\undefined}
}%
\providecommand \@ifnum [1]{%
 \ifnum #1\expandafter \@firstoftwo
 \else \expandafter \@secondoftwo
 \fi
}%
\providecommand \@ifx [1]{%
 \ifx #1\expandafter \@firstoftwo
 \else \expandafter \@secondoftwo
 \fi
}%
\providecommand \natexlab [1]{#1}%
\providecommand \enquote  [1]{``#1''}%
\providecommand \bibnamefont  [1]{#1}%
\providecommand \bibfnamefont [1]{#1}%
\providecommand \citenamefont [1]{#1}%
\providecommand \href@noop [0]{\@secondoftwo}%
\providecommand \href [0]{\begingroup \@sanitize@url \@href}%
\providecommand \@href[1]{\@@startlink{#1}\@@href}%
\providecommand \@@href[1]{\endgroup#1\@@endlink}%
\providecommand \@sanitize@url [0]{\catcode `\\12\catcode `\$12\catcode `\&12\catcode `\#12\catcode `\^12\catcode `\_12\catcode `\%12\relax}%
\providecommand \@@startlink[1]{}%
\providecommand \@@endlink[0]{}%
\providecommand \url  [0]{\begingroup\@sanitize@url \@url }%
\providecommand \@url [1]{\endgroup\@href {#1}{\urlprefix }}%
\providecommand \urlprefix  [0]{URL }%
\providecommand \Eprint [0]{\href }%
\providecommand \doibase [0]{https://doi.org/}%
\providecommand \selectlanguage [0]{\@gobble}%
\providecommand \bibinfo  [0]{\@secondoftwo}%
\providecommand \bibfield  [0]{\@secondoftwo}%
\providecommand \translation [1]{[#1]}%
\providecommand \BibitemOpen [0]{}%
\providecommand \bibitemStop [0]{}%
\providecommand \bibitemNoStop [0]{.\EOS\space}%
\providecommand \EOS [0]{\spacefactor3000\relax}%
\providecommand \BibitemShut  [1]{\csname bibitem#1\endcsname}%
\let\auto@bib@innerbib\@empty
\bibitem [{\citenamefont {Shinjo}\ \emph {et~al.}(2000)\citenamefont {Shinjo}, \citenamefont {Okuno}, \citenamefont {Hassdorf}, \citenamefont {Shigeto},\ and\ \citenamefont {Ono}}]{Shinjo2000}%
  \BibitemOpen
  \bibfield  {author} {\bibinfo {author} {\bibfnamefont {T.}~\bibnamefont {Shinjo}}, \bibinfo {author} {\bibfnamefont {T.}~\bibnamefont {Okuno}}, \bibinfo {author} {\bibfnamefont {R.}~\bibnamefont {Hassdorf}}, \bibinfo {author} {\bibfnamefont {K.}~\bibnamefont {Shigeto}},\ and\ \bibinfo {author} {\bibfnamefont {T.}~\bibnamefont {Ono}},\ }\bibfield  {title} {\bibinfo {title} {Magnetic vortex core observation in circular dots of permalloy},\ }\href {https://doi.org/10.1126/science.289.5481.930} {\bibfield  {journal} {\bibinfo  {journal} {Science}\ }\textbf {\bibinfo {volume} {289}},\ \bibinfo {pages} {930} (\bibinfo {year} {2000})}\BibitemShut {NoStop}%
\bibitem [{\citenamefont {Choe}\ \emph {et~al.}(2004)\citenamefont {Choe}, \citenamefont {Acremann}, \citenamefont {Scholl}, \citenamefont {Bauer}, \citenamefont {Doran}, \citenamefont {Stöhr},\ and\ \citenamefont {Padmore}}]{Choe2004}%
  \BibitemOpen
  \bibfield  {author} {\bibinfo {author} {\bibfnamefont {S.-B.}\ \bibnamefont {Choe}}, \bibinfo {author} {\bibfnamefont {Y.}~\bibnamefont {Acremann}}, \bibinfo {author} {\bibfnamefont {A.}~\bibnamefont {Scholl}}, \bibinfo {author} {\bibfnamefont {A.}~\bibnamefont {Bauer}}, \bibinfo {author} {\bibfnamefont {A.}~\bibnamefont {Doran}}, \bibinfo {author} {\bibfnamefont {J.}~\bibnamefont {Stöhr}},\ and\ \bibinfo {author} {\bibfnamefont {H.~A.}\ \bibnamefont {Padmore}},\ }\bibfield  {title} {\bibinfo {title} {Vortex core-driven magnetization dynamics},\ }\href {https://doi.org/10.1126/science.1095068} {\bibfield  {journal} {\bibinfo  {journal} {Science}\ }\textbf {\bibinfo {volume} {304}},\ \bibinfo {pages} {420} (\bibinfo {year} {2004})}\BibitemShut {NoStop}%
\bibitem [{\citenamefont {Guslienko}\ \emph {et~al.}(2006)\citenamefont {Guslienko}, \citenamefont {Han}, \citenamefont {Keavney}, \citenamefont {Divan},\ and\ \citenamefont {Bader}}]{Guslienko2006}%
  \BibitemOpen
  \bibfield  {author} {\bibinfo {author} {\bibfnamefont {K.~Y.}\ \bibnamefont {Guslienko}}, \bibinfo {author} {\bibfnamefont {X.~F.}\ \bibnamefont {Han}}, \bibinfo {author} {\bibfnamefont {D.~J.}\ \bibnamefont {Keavney}}, \bibinfo {author} {\bibfnamefont {R.}~\bibnamefont {Divan}},\ and\ \bibinfo {author} {\bibfnamefont {S.~D.}\ \bibnamefont {Bader}},\ }\bibfield  {title} {\bibinfo {title} {Magnetic vortex core dynamics in cylindrical ferromagnetic dots},\ }\href {https://doi.org/10.1103/PhysRevLett.96.067205} {\bibfield  {journal} {\bibinfo  {journal} {Phys. Rev. Lett.}\ }\textbf {\bibinfo {volume} {96}},\ \bibinfo {pages} {067205} (\bibinfo {year} {2006})}\BibitemShut {NoStop}%
\bibitem [{\citenamefont {Pribiag}\ \emph {et~al.}(2007)\citenamefont {Pribiag}, \citenamefont {Krivorotov}, \citenamefont {Fuchs}, \citenamefont {Braganca}, \citenamefont {Ozatay}, \citenamefont {Sankey}, \citenamefont {Ralph},\ and\ \citenamefont {Buhrman}}]{Pribiag2007}%
  \BibitemOpen
  \bibfield  {author} {\bibinfo {author} {\bibfnamefont {V.~S.}\ \bibnamefont {Pribiag}}, \bibinfo {author} {\bibfnamefont {I.~N.}\ \bibnamefont {Krivorotov}}, \bibinfo {author} {\bibfnamefont {G.~D.}\ \bibnamefont {Fuchs}}, \bibinfo {author} {\bibfnamefont {P.~M.}\ \bibnamefont {Braganca}}, \bibinfo {author} {\bibfnamefont {O.}~\bibnamefont {Ozatay}}, \bibinfo {author} {\bibfnamefont {J.~C.}\ \bibnamefont {Sankey}}, \bibinfo {author} {\bibfnamefont {D.~C.}\ \bibnamefont {Ralph}},\ and\ \bibinfo {author} {\bibfnamefont {R.~A.}\ \bibnamefont {Buhrman}},\ }\bibfield  {title} {\bibinfo {title} {Magnetic vortex oscillator driven by d.c. spin-polarized current},\ }\href {https://doi.org/10.1038/nphys619} {\bibfield  {journal} {\bibinfo  {journal} {Nat. Phys.}\ }\textbf {\bibinfo {volume} {3}},\ \bibinfo {pages} {498} (\bibinfo {year} {2007})}\BibitemShut {NoStop}%
\bibitem [{\citenamefont {Mistral}\ \emph {et~al.}(2008)\citenamefont {Mistral}, \citenamefont {van Kampen}, \citenamefont {Hrkac}, \citenamefont {Kim}, \citenamefont {Devolder}, \citenamefont {Crozat}, \citenamefont {Chappert}, \citenamefont {Lagae},\ and\ \citenamefont {Schrefl}}]{Mistral2008}%
  \BibitemOpen
  \bibfield  {author} {\bibinfo {author} {\bibfnamefont {Q.}~\bibnamefont {Mistral}}, \bibinfo {author} {\bibfnamefont {M.}~\bibnamefont {van Kampen}}, \bibinfo {author} {\bibfnamefont {G.}~\bibnamefont {Hrkac}}, \bibinfo {author} {\bibfnamefont {J.-V.}\ \bibnamefont {Kim}}, \bibinfo {author} {\bibfnamefont {T.}~\bibnamefont {Devolder}}, \bibinfo {author} {\bibfnamefont {P.}~\bibnamefont {Crozat}}, \bibinfo {author} {\bibfnamefont {C.}~\bibnamefont {Chappert}}, \bibinfo {author} {\bibfnamefont {L.}~\bibnamefont {Lagae}},\ and\ \bibinfo {author} {\bibfnamefont {T.}~\bibnamefont {Schrefl}},\ }\bibfield  {title} {\bibinfo {title} {Current-driven vortex oscillations in metallic nanocontacts},\ }\href {https://doi.org/10.1103/PhysRevLett.100.257201} {\bibfield  {journal} {\bibinfo  {journal} {Phys. Rev. Lett.}\ }\textbf {\bibinfo {volume} {100}},\ \bibinfo {pages} {257201} (\bibinfo {year} {2008})}\BibitemShut {NoStop}%
\bibitem [{\citenamefont {Dussaux}\ \emph {et~al.}(2010)\citenamefont {Dussaux}, \citenamefont {Georges}, \citenamefont {Grollier}, \citenamefont {Cros}, \citenamefont {Khvalkovskiy}, \citenamefont {Fukushima}, \citenamefont {Konoto}, \citenamefont {Kubota}, \citenamefont {Yakushiji}, \citenamefont {Yuasa}, \citenamefont {Zvezdin}, \citenamefont {Ando},\ and\ \citenamefont {Fert}}]{Dussaux2010}%
  \BibitemOpen
  \bibfield  {author} {\bibinfo {author} {\bibfnamefont {A.}~\bibnamefont {Dussaux}}, \bibinfo {author} {\bibfnamefont {B.}~\bibnamefont {Georges}}, \bibinfo {author} {\bibfnamefont {J.}~\bibnamefont {Grollier}}, \bibinfo {author} {\bibfnamefont {V.}~\bibnamefont {Cros}}, \bibinfo {author} {\bibfnamefont {A.~V.}\ \bibnamefont {Khvalkovskiy}}, \bibinfo {author} {\bibfnamefont {A.}~\bibnamefont {Fukushima}}, \bibinfo {author} {\bibfnamefont {M.}~\bibnamefont {Konoto}}, \bibinfo {author} {\bibfnamefont {H.}~\bibnamefont {Kubota}}, \bibinfo {author} {\bibfnamefont {K.}~\bibnamefont {Yakushiji}}, \bibinfo {author} {\bibfnamefont {S.}~\bibnamefont {Yuasa}}, \bibinfo {author} {\bibfnamefont {K.~A.}\ \bibnamefont {Zvezdin}}, \bibinfo {author} {\bibfnamefont {K.}~\bibnamefont {Ando}},\ and\ \bibinfo {author} {\bibfnamefont {A.}~\bibnamefont {Fert}},\ }\bibfield  {title} {\bibinfo {title} {Large microwave generation from current-driven magnetic vortex oscillators in magnetic tunnel junctions},\ }\href {https://doi.org/10.1038/ncomms1006} {\bibfield  {journal} {\bibinfo  {journal} {Nat. Commun.}\ }\textbf {\bibinfo {volume} {1}},\ \bibinfo {pages} {8} (\bibinfo {year} {2010})}\BibitemShut {NoStop}%
\bibitem [{\citenamefont {Torrejon}\ \emph {et~al.}(2017)\citenamefont {Torrejon}, \citenamefont {Riou}, \citenamefont {Abreu~Araujo}, \citenamefont {Tsunegi}, \citenamefont {Khalsa}, \citenamefont {Querlioz}, \citenamefont {Bortolotti}, \citenamefont {Cros}, \citenamefont {Yakushiji}, \citenamefont {Fukushima}, \citenamefont {Kubota}, \citenamefont {Yuasa}, \citenamefont {Stiles},\ and\ \citenamefont {Grollier}}]{Torrejon2017}%
  \BibitemOpen
  \bibfield  {author} {\bibinfo {author} {\bibfnamefont {J.}~\bibnamefont {Torrejon}}, \bibinfo {author} {\bibfnamefont {M.}~\bibnamefont {Riou}}, \bibinfo {author} {\bibfnamefont {F.}~\bibnamefont {Abreu~Araujo}}, \bibinfo {author} {\bibfnamefont {S.}~\bibnamefont {Tsunegi}}, \bibinfo {author} {\bibfnamefont {G.}~\bibnamefont {Khalsa}}, \bibinfo {author} {\bibfnamefont {D.}~\bibnamefont {Querlioz}}, \bibinfo {author} {\bibfnamefont {P.}~\bibnamefont {Bortolotti}}, \bibinfo {author} {\bibfnamefont {V.}~\bibnamefont {Cros}}, \bibinfo {author} {\bibfnamefont {K.}~\bibnamefont {Yakushiji}}, \bibinfo {author} {\bibfnamefont {A.}~\bibnamefont {Fukushima}}, \bibinfo {author} {\bibfnamefont {H.}~\bibnamefont {Kubota}}, \bibinfo {author} {\bibfnamefont {S.}~\bibnamefont {Yuasa}}, \bibinfo {author} {\bibfnamefont {M.~D.}\ \bibnamefont {Stiles}},\ and\ \bibinfo {author} {\bibfnamefont {J.}~\bibnamefont {Grollier}},\ }\bibfield  {title} {\bibinfo {title} {Neuromorphic computing with nanoscale spintronic oscillators},\ }\href {https://doi.org/10.1038/nature23011} {\bibfield  {journal} {\bibinfo  {journal} {Nature}\ }\textbf {\bibinfo {volume} {547}},\ \bibinfo {pages} {428} (\bibinfo {year} {2017})}\BibitemShut {NoStop}%
\bibitem [{\citenamefont {Romera}\ \emph {et~al.}(2018)\citenamefont {Romera}, \citenamefont {Talatchian}, \citenamefont {Tsunegi}, \citenamefont {Abreu~Araujo}, \citenamefont {Cros}, \citenamefont {Bortolotti}, \citenamefont {Trastoy}, \citenamefont {Yakushiji}, \citenamefont {Fukushima}, \citenamefont {Kubota}, \citenamefont {Yuasa}, \citenamefont {Ernoult}, \citenamefont {Vodenicarevic}, \citenamefont {Hirtzlin}, \citenamefont {Locatelli}, \citenamefont {Querlioz},\ and\ \citenamefont {Grollier}}]{Romera2018}%
  \BibitemOpen
  \bibfield  {author} {\bibinfo {author} {\bibfnamefont {M.}~\bibnamefont {Romera}}, \bibinfo {author} {\bibfnamefont {P.}~\bibnamefont {Talatchian}}, \bibinfo {author} {\bibfnamefont {S.}~\bibnamefont {Tsunegi}}, \bibinfo {author} {\bibfnamefont {F.}~\bibnamefont {Abreu~Araujo}}, \bibinfo {author} {\bibfnamefont {V.}~\bibnamefont {Cros}}, \bibinfo {author} {\bibfnamefont {P.}~\bibnamefont {Bortolotti}}, \bibinfo {author} {\bibfnamefont {J.}~\bibnamefont {Trastoy}}, \bibinfo {author} {\bibfnamefont {K.}~\bibnamefont {Yakushiji}}, \bibinfo {author} {\bibfnamefont {A.}~\bibnamefont {Fukushima}}, \bibinfo {author} {\bibfnamefont {H.}~\bibnamefont {Kubota}}, \bibinfo {author} {\bibfnamefont {S.}~\bibnamefont {Yuasa}}, \bibinfo {author} {\bibfnamefont {M.}~\bibnamefont {Ernoult}}, \bibinfo {author} {\bibfnamefont {D.}~\bibnamefont {Vodenicarevic}}, \bibinfo {author} {\bibfnamefont {T.}~\bibnamefont {Hirtzlin}}, \bibinfo {author} {\bibfnamefont {N.}~\bibnamefont {Locatelli}}, \bibinfo {author} {\bibfnamefont {D.}~\bibnamefont {Querlioz}},\ and\ \bibinfo {author} {\bibfnamefont {J.}~\bibnamefont {Grollier}},\ }\bibfield  {title} {\bibinfo {title} {Vowel recognition with four coupled spin-torque nano-oscillators},\ }\href {https://doi.org/10.1038/s41586-018-0632-y} {\bibfield  {journal} {\bibinfo  {journal} {Nature}\ }\textbf {\bibinfo {volume} {563}},\ \bibinfo {pages} {230} (\bibinfo {year} {2018})}\BibitemShut {NoStop}%
\bibitem [{\citenamefont {Slavin}\ and\ \citenamefont {Tiberkevich}(2009)}]{Slavin2009}%
  \BibitemOpen
  \bibfield  {author} {\bibinfo {author} {\bibfnamefont {A.~N.}\ \bibnamefont {Slavin}}\ and\ \bibinfo {author} {\bibfnamefont {V.~S.}\ \bibnamefont {Tiberkevich}},\ }\bibfield  {title} {\bibinfo {title} {Nonlinear auto-oscillator theory of microwave generation by spin-polarized current},\ }\href {https://doi.org/10.1109/TMAG.2008.2009935} {\bibfield  {journal} {\bibinfo  {journal} {IEEE Trans. Magn.}\ }\textbf {\bibinfo {volume} {45}},\ \bibinfo {pages} {1875} (\bibinfo {year} {2009})}\BibitemShut {NoStop}%
\bibitem [{\citenamefont {Hamadeh}\ \emph {et~al.}(2014)\citenamefont {Hamadeh}, \citenamefont {Locatelli}, \citenamefont {Naletov}, \citenamefont {Lebrun}, \citenamefont {de~Loubens}, \citenamefont {Grollier}, \citenamefont {Klein},\ and\ \citenamefont {Cros}}]{Hamadeh2014}%
  \BibitemOpen
  \bibfield  {author} {\bibinfo {author} {\bibfnamefont {A.}~\bibnamefont {Hamadeh}}, \bibinfo {author} {\bibfnamefont {N.}~\bibnamefont {Locatelli}}, \bibinfo {author} {\bibfnamefont {V.~V.}\ \bibnamefont {Naletov}}, \bibinfo {author} {\bibfnamefont {R.}~\bibnamefont {Lebrun}}, \bibinfo {author} {\bibfnamefont {G.}~\bibnamefont {de~Loubens}}, \bibinfo {author} {\bibfnamefont {J.}~\bibnamefont {Grollier}}, \bibinfo {author} {\bibfnamefont {O.}~\bibnamefont {Klein}},\ and\ \bibinfo {author} {\bibfnamefont {V.}~\bibnamefont {Cros}},\ }\bibfield  {title} {\bibinfo {title} {Origin of spectral purity and tuning sensitivity in a spin transfer vortex nano-oscillator},\ }\href {https://doi.org/10.1103/PhysRevLett.112.257201} {\bibfield  {journal} {\bibinfo  {journal} {Phys. Rev. Lett.}\ }\textbf {\bibinfo {volume} {112}},\ \bibinfo {pages} {257201} (\bibinfo {year} {2014})}\BibitemShut {NoStop}%
\bibitem [{\citenamefont {Thiele}(1973)}]{Thiele1973}%
  \BibitemOpen
  \bibfield  {author} {\bibinfo {author} {\bibfnamefont {A.~A.}\ \bibnamefont {Thiele}},\ }\bibfield  {title} {\bibinfo {title} {Steady-state motion of magnetic domains},\ }\href {https://doi.org/10.1103/PhysRevLett.30.230} {\bibfield  {journal} {\bibinfo  {journal} {Phys. Rev. Lett.}\ }\textbf {\bibinfo {volume} {30}},\ \bibinfo {pages} {230} (\bibinfo {year} {1973})}\BibitemShut {NoStop}%
\bibitem [{\citenamefont {Huber}(1982)}]{Huber1982}%
  \BibitemOpen
  \bibfield  {author} {\bibinfo {author} {\bibfnamefont {D.~L.}\ \bibnamefont {Huber}},\ }\bibfield  {title} {\bibinfo {title} {Dynamics of spin vortices in two-dimensional planar magnets},\ }\href {https://doi.org/10.1103/PhysRevB.26.3758} {\bibfield  {journal} {\bibinfo  {journal} {Phys. Rev. B}\ }\textbf {\bibinfo {volume} {26}},\ \bibinfo {pages} {3758} (\bibinfo {year} {1982})}\BibitemShut {NoStop}%
\bibitem [{\citenamefont {Tretiakov}\ \emph {et~al.}(2008)\citenamefont {Tretiakov}, \citenamefont {Clarke}, \citenamefont {Chern}, \citenamefont {Bazaliy},\ and\ \citenamefont {Tchernyshyov}}]{Tretiakov2008}%
  \BibitemOpen
  \bibfield  {author} {\bibinfo {author} {\bibfnamefont {O.~A.}\ \bibnamefont {Tretiakov}}, \bibinfo {author} {\bibfnamefont {D.}~\bibnamefont {Clarke}}, \bibinfo {author} {\bibfnamefont {G.-W.}\ \bibnamefont {Chern}}, \bibinfo {author} {\bibfnamefont {Y.~B.}\ \bibnamefont {Bazaliy}},\ and\ \bibinfo {author} {\bibfnamefont {O.}~\bibnamefont {Tchernyshyov}},\ }\bibfield  {title} {\bibinfo {title} {Dynamics of domain walls in magnetic nanostrips},\ }\href {https://doi.org/10.1103/PhysRevLett.100.127204} {\bibfield  {journal} {\bibinfo  {journal} {Phys. Rev. Lett.}\ }\textbf {\bibinfo {volume} {100}},\ \bibinfo {pages} {127204} (\bibinfo {year} {2008})}\BibitemShut {NoStop}%
\bibitem [{\citenamefont {Metlov}(2013)}]{Metlov2013}%
  \BibitemOpen
  \bibfield  {author} {\bibinfo {author} {\bibfnamefont {K.~L.}\ \bibnamefont {Metlov}},\ }\bibfield  {title} {\bibinfo {title} {Vortex mechanics in planar nanomagnets},\ }\href {https://doi.org/10.1103/PhysRevB.88.014427} {\bibfield  {journal} {\bibinfo  {journal} {Phys. Rev. B}\ }\textbf {\bibinfo {volume} {88}},\ \bibinfo {pages} {014427} (\bibinfo {year} {2013})}\BibitemShut {NoStop}%
\bibitem [{\citenamefont {D{\"o}ring}(1948)}]{Doring1948}%
  \BibitemOpen
  \bibfield  {author} {\bibinfo {author} {\bibfnamefont {W.}~\bibnamefont {D{\"o}ring}},\ }\bibfield  {title} {\bibinfo {title} {{\"U}ber die tr{\"a}gheit der w{\"a}nde zwischen wei{\ss}schen bezirken},\ }\href {https://doi.org/10.1515/zna-1948-0701} {\bibfield  {journal} {\bibinfo  {journal} {Z. Naturforsch. A}\ }\textbf {\bibinfo {volume} {3}},\ \bibinfo {pages} {373} (\bibinfo {year} {1948})}\BibitemShut {NoStop}%
\bibitem [{\citenamefont {Gilbert}(2004)}]{Gilbert2004}%
  \BibitemOpen
  \bibfield  {author} {\bibinfo {author} {\bibfnamefont {T.~L.}\ \bibnamefont {Gilbert}},\ }\bibfield  {title} {\bibinfo {title} {A phenomenological theory of damping in ferromagnetic materials},\ }\href {https://doi.org/10.1109/TMAG.2004.836740} {\bibfield  {journal} {\bibinfo  {journal} {IEEE Trans. Magn.}\ }\textbf {\bibinfo {volume} {40}},\ \bibinfo {pages} {3443} (\bibinfo {year} {2004})}\BibitemShut {NoStop}%
\bibitem [{\citenamefont {Papanicolaou}\ and\ \citenamefont {Tomaras}(1991)}]{Papanicolaou1991}%
  \BibitemOpen
  \bibfield  {author} {\bibinfo {author} {\bibfnamefont {N.}~\bibnamefont {Papanicolaou}}\ and\ \bibinfo {author} {\bibfnamefont {T.~N.}\ \bibnamefont {Tomaras}},\ }\bibfield  {title} {\bibinfo {title} {Dynamics of magnetic vortices},\ }\href {https://doi.org/10.1016/0550-3213(91)90410-Y} {\bibfield  {journal} {\bibinfo  {journal} {Nucl. Phys. B}\ }\textbf {\bibinfo {volume} {360}},\ \bibinfo {pages} {425} (\bibinfo {year} {1991})}\BibitemShut {NoStop}%
\bibitem [{\citenamefont {Guslienko}\ \emph {et~al.}(2002)\citenamefont {Guslienko}, \citenamefont {Ivanov}, \citenamefont {Novosad}, \citenamefont {Otani}, \citenamefont {Shima},\ and\ \citenamefont {Fukamichi}}]{Guslienko2002}%
  \BibitemOpen
  \bibfield  {author} {\bibinfo {author} {\bibfnamefont {K.~Y.}\ \bibnamefont {Guslienko}}, \bibinfo {author} {\bibfnamefont {B.~A.}\ \bibnamefont {Ivanov}}, \bibinfo {author} {\bibfnamefont {V.}~\bibnamefont {Novosad}}, \bibinfo {author} {\bibfnamefont {Y.}~\bibnamefont {Otani}}, \bibinfo {author} {\bibfnamefont {H.}~\bibnamefont {Shima}},\ and\ \bibinfo {author} {\bibfnamefont {K.}~\bibnamefont {Fukamichi}},\ }\bibfield  {title} {\bibinfo {title} {Eigenfrequencies of vortex state excitations in magnetic submicron-size disks},\ }\href {https://doi.org/10.1063/1.1450816} {\bibfield  {journal} {\bibinfo  {journal} {J. Appl. Phys.}\ }\textbf {\bibinfo {volume} {91}},\ \bibinfo {pages} {8037} (\bibinfo {year} {2002})}\BibitemShut {NoStop}%
\bibitem [{\citenamefont {Antos}\ \emph {et~al.}(2008)\citenamefont {Antos}, \citenamefont {Otani},\ and\ \citenamefont {Shibata}}]{Antos2008}%
  \BibitemOpen
  \bibfield  {author} {\bibinfo {author} {\bibfnamefont {R.}~\bibnamefont {Antos}}, \bibinfo {author} {\bibfnamefont {Y.}~\bibnamefont {Otani}},\ and\ \bibinfo {author} {\bibfnamefont {J.}~\bibnamefont {Shibata}},\ }\bibfield  {title} {\bibinfo {title} {Magnetic vortex dynamics},\ }\href {https://doi.org/10.1143/JPSJ.77.031004} {\bibfield  {journal} {\bibinfo  {journal} {J. Phys. Soc. Jpn.}\ }\textbf {\bibinfo {volume} {77}},\ \bibinfo {pages} {031004} (\bibinfo {year} {2008})}\BibitemShut {NoStop}%
\bibitem [{\citenamefont {Guslienko}(2006)}]{Guslienko2006_APL}%
  \BibitemOpen
  \bibfield  {author} {\bibinfo {author} {\bibfnamefont {K.~Y.}\ \bibnamefont {Guslienko}},\ }\bibfield  {title} {\bibinfo {title} {Low-frequency vortex dynamic susceptibility and relaxation in mesoscopic ferromagnetic dots},\ }\href {https://doi.org/10.1063/1.2221904} {\bibfield  {journal} {\bibinfo  {journal} {Appl. Phys. Lett.}\ }\textbf {\bibinfo {volume} {89}},\ \bibinfo {pages} {022510} (\bibinfo {year} {2006})}\BibitemShut {NoStop}%
\bibitem [{\citenamefont {Metlov}\ and\ \citenamefont {Guslienko}(2002)}]{Metlov2002}%
  \BibitemOpen
  \bibfield  {author} {\bibinfo {author} {\bibfnamefont {K.~L.}\ \bibnamefont {Metlov}}\ and\ \bibinfo {author} {\bibfnamefont {K.~Y.}\ \bibnamefont {Guslienko}},\ }\bibfield  {title} {\bibinfo {title} {Stability of magnetic vortex in soft magnetic nano-sized circular cylinder},\ }\href {https://doi.org/10.1016/S0304-8853(01)01360-9} {\bibfield  {journal} {\bibinfo  {journal} {J. Magn. Magn. Mater.}\ }\textbf {\bibinfo {volume} {242--245}},\ \bibinfo {pages} {1015} (\bibinfo {year} {2002})}\BibitemShut {NoStop}%
\bibitem [{\citenamefont {de~Wergifosse}\ \emph {et~al.}(2023)\citenamefont {de~Wergifosse}, \citenamefont {Chopin},\ and\ \citenamefont {Abreu~Araujo}}]{deWergifosse2023}%
  \BibitemOpen
  \bibfield  {author} {\bibinfo {author} {\bibfnamefont {S.}~\bibnamefont {de~Wergifosse}}, \bibinfo {author} {\bibfnamefont {C.}~\bibnamefont {Chopin}},\ and\ \bibinfo {author} {\bibfnamefont {F.}~\bibnamefont {Abreu~Araujo}},\ }\bibfield  {title} {\bibinfo {title} {Quantitative and realistic description of the magnetic potential energy of spin-torque vortex oscillators},\ }\href {https://doi.org/10.1103/PhysRevB.108.174403} {\bibfield  {journal} {\bibinfo  {journal} {Phys. Rev. B}\ }\textbf {\bibinfo {volume} {108}},\ \bibinfo {pages} {174403} (\bibinfo {year} {2023})}\BibitemShut {NoStop}%
\bibitem [{\citenamefont {Vansteenkiste}\ \emph {et~al.}(2009)\citenamefont {Vansteenkiste}, \citenamefont {Chou}, \citenamefont {Weigand}, \citenamefont {Curcic}, \citenamefont {Sackmann}, \citenamefont {Stoll}, \citenamefont {Tyliszczak}, \citenamefont {Woltersdorf}, \citenamefont {Back}, \citenamefont {Sch{\"u}tz},\ and\ \citenamefont {Van~Waeyenberge}}]{Vansteenkiste2009}%
  \BibitemOpen
  \bibfield  {author} {\bibinfo {author} {\bibfnamefont {A.}~\bibnamefont {Vansteenkiste}}, \bibinfo {author} {\bibfnamefont {K.~W.}\ \bibnamefont {Chou}}, \bibinfo {author} {\bibfnamefont {M.}~\bibnamefont {Weigand}}, \bibinfo {author} {\bibfnamefont {M.}~\bibnamefont {Curcic}}, \bibinfo {author} {\bibfnamefont {V.}~\bibnamefont {Sackmann}}, \bibinfo {author} {\bibfnamefont {H.}~\bibnamefont {Stoll}}, \bibinfo {author} {\bibfnamefont {T.}~\bibnamefont {Tyliszczak}}, \bibinfo {author} {\bibfnamefont {G.}~\bibnamefont {Woltersdorf}}, \bibinfo {author} {\bibfnamefont {C.~H.}\ \bibnamefont {Back}}, \bibinfo {author} {\bibfnamefont {G.}~\bibnamefont {Sch{\"u}tz}},\ and\ \bibinfo {author} {\bibfnamefont {B.}~\bibnamefont {Van~Waeyenberge}},\ }\bibfield  {title} {\bibinfo {title} {X-ray imaging of the dynamic magnetic vortex core deformation},\ }\href {https://doi.org/10.1038/nphys1231} {\bibfield  {journal} {\bibinfo  {journal} {Nat. Phys.}\ }\textbf {\bibinfo {volume} {5}},\ \bibinfo {pages} {332} (\bibinfo {year} {2009})}\BibitemShut {NoStop}%
\bibitem [{\citenamefont {Dussaux}\ \emph {et~al.}(2012)\citenamefont {Dussaux}, \citenamefont {Khvalkovskiy}, \citenamefont {Bortolotti}, \citenamefont {Grollier}, \citenamefont {Cros},\ and\ \citenamefont {Fert}}]{Dussaux2012}%
  \BibitemOpen
  \bibfield  {author} {\bibinfo {author} {\bibfnamefont {A.}~\bibnamefont {Dussaux}}, \bibinfo {author} {\bibfnamefont {A.~V.}\ \bibnamefont {Khvalkovskiy}}, \bibinfo {author} {\bibfnamefont {P.}~\bibnamefont {Bortolotti}}, \bibinfo {author} {\bibfnamefont {J.}~\bibnamefont {Grollier}}, \bibinfo {author} {\bibfnamefont {V.}~\bibnamefont {Cros}},\ and\ \bibinfo {author} {\bibfnamefont {A.}~\bibnamefont {Fert}},\ }\bibfield  {title} {\bibinfo {title} {Field dependence of spin-transfer-induced vortex dynamics in the nonlinear regime},\ }\href {https://doi.org/10.1103/PhysRevB.86.014402} {\bibfield  {journal} {\bibinfo  {journal} {Phys. Rev. B}\ }\textbf {\bibinfo {volume} {86}},\ \bibinfo {pages} {014402} (\bibinfo {year} {2012})}\BibitemShut {NoStop}%
\bibitem [{\citenamefont {Yoo}\ \emph {et~al.}(2021)\citenamefont {Yoo}, \citenamefont {Mineo},\ and\ \citenamefont {Kim}}]{Yoo2021}%
  \BibitemOpen
  \bibfield  {author} {\bibinfo {author} {\bibfnamefont {M.-W.}\ \bibnamefont {Yoo}}, \bibinfo {author} {\bibfnamefont {F.}~\bibnamefont {Mineo}},\ and\ \bibinfo {author} {\bibfnamefont {J.-V.}\ \bibnamefont {Kim}},\ }\bibfield  {title} {\bibinfo {title} {Analytical model of the deformation-induced inertial dynamics of a magnetic vortex},\ }\href {https://doi.org/10.1063/5.0039535} {\bibfield  {journal} {\bibinfo  {journal} {J. Appl. Phys.}\ }\textbf {\bibinfo {volume} {129}},\ \bibinfo {pages} {053903} (\bibinfo {year} {2021})}\BibitemShut {NoStop}%
\bibitem [{\citenamefont {Lee}\ and\ \citenamefont {Kim}(2007)}]{Lee2007}%
  \BibitemOpen
  \bibfield  {author} {\bibinfo {author} {\bibfnamefont {K.-S.}\ \bibnamefont {Lee}}\ and\ \bibinfo {author} {\bibfnamefont {S.-K.}\ \bibnamefont {Kim}},\ }\bibfield  {title} {\bibinfo {title} {Gyrotropic linear and nonlinear motions of a magnetic vortex in soft magnetic nanodots},\ }\href {https://doi.org/10.1063/1.2783272} {\bibfield  {journal} {\bibinfo  {journal} {Appl. Phys. Lett.}\ }\textbf {\bibinfo {volume} {91}},\ \bibinfo {pages} {132511} (\bibinfo {year} {2007})}\BibitemShut {NoStop}%
\bibitem [{\citenamefont {Khvalkovskiy}\ \emph {et~al.}(2009)\citenamefont {Khvalkovskiy}, \citenamefont {Grollier}, \citenamefont {Dussaux}, \citenamefont {Zvezdin},\ and\ \citenamefont {Cros}}]{Khvalkovskiy2009}%
  \BibitemOpen
  \bibfield  {author} {\bibinfo {author} {\bibfnamefont {A.~V.}\ \bibnamefont {Khvalkovskiy}}, \bibinfo {author} {\bibfnamefont {J.}~\bibnamefont {Grollier}}, \bibinfo {author} {\bibfnamefont {A.}~\bibnamefont {Dussaux}}, \bibinfo {author} {\bibfnamefont {K.~A.}\ \bibnamefont {Zvezdin}},\ and\ \bibinfo {author} {\bibfnamefont {V.}~\bibnamefont {Cros}},\ }\bibfield  {title} {\bibinfo {title} {Vortex oscillations induced by spin-polarized current in a magnetic nanopillar: Analytical versus micromagnetic calculations},\ }\href {https://doi.org/10.1103/PhysRevB.80.140401} {\bibfield  {journal} {\bibinfo  {journal} {Phys. Rev. B}\ }\textbf {\bibinfo {volume} {80}},\ \bibinfo {pages} {140401} (\bibinfo {year} {2009})}\BibitemShut {NoStop}%
\bibitem [{\citenamefont {Guslienko}\ \emph {et~al.}(2010)\citenamefont {Guslienko}, \citenamefont {Hern{\'a}ndez~Heredero},\ and\ \citenamefont {Chubykalo-Fesenko}}]{Guslienko2010Nonlinear}%
  \BibitemOpen
  \bibfield  {author} {\bibinfo {author} {\bibfnamefont {K.~Y.}\ \bibnamefont {Guslienko}}, \bibinfo {author} {\bibfnamefont {R.}~\bibnamefont {Hern{\'a}ndez~Heredero}},\ and\ \bibinfo {author} {\bibfnamefont {O.}~\bibnamefont {Chubykalo-Fesenko}},\ }\bibfield  {title} {\bibinfo {title} {Nonlinear gyrotropic vortex dynamics in ferromagnetic dots},\ }\href {https://doi.org/10.1103/PhysRevB.82.014402} {\bibfield  {journal} {\bibinfo  {journal} {Phys. Rev. B}\ }\textbf {\bibinfo {volume} {82}},\ \bibinfo {pages} {014402} (\bibinfo {year} {2010})}\BibitemShut {NoStop}%
\bibitem [{\citenamefont {Drews}\ \emph {et~al.}(2012)\citenamefont {Drews}, \citenamefont {Krüger}, \citenamefont {Selke}, \citenamefont {Kamionka}, \citenamefont {Vogel}, \citenamefont {Martens}, \citenamefont {Merkt}, \citenamefont {Möller},\ and\ \citenamefont {Meier}}]{Drews2012}%
  \BibitemOpen
  \bibfield  {author} {\bibinfo {author} {\bibfnamefont {A.}~\bibnamefont {Drews}}, \bibinfo {author} {\bibfnamefont {B.}~\bibnamefont {Krüger}}, \bibinfo {author} {\bibfnamefont {G.}~\bibnamefont {Selke}}, \bibinfo {author} {\bibfnamefont {T.}~\bibnamefont {Kamionka}}, \bibinfo {author} {\bibfnamefont {A.}~\bibnamefont {Vogel}}, \bibinfo {author} {\bibfnamefont {M.}~\bibnamefont {Martens}}, \bibinfo {author} {\bibfnamefont {U.}~\bibnamefont {Merkt}}, \bibinfo {author} {\bibfnamefont {D.}~\bibnamefont {Möller}},\ and\ \bibinfo {author} {\bibfnamefont {G.}~\bibnamefont {Meier}},\ }\bibfield  {title} {\bibinfo {title} {Nonlinear magnetic vortex gyration},\ }\href {https://doi.org/10.1103/PhysRevB.85.144417} {\bibfield  {journal} {\bibinfo  {journal} {Phys. Rev. B}\ }\textbf {\bibinfo {volume} {85}},\ \bibinfo {pages} {144417} (\bibinfo {year} {2012})}\BibitemShut {NoStop}%
\bibitem [{\citenamefont {Shim}\ \emph {et~al.}(2011)\citenamefont {Shim}, \citenamefont {Piao}, \citenamefont {Lee}, \citenamefont {Oh}, \citenamefont {Yu}, \citenamefont {Han},\ and\ \citenamefont {Kim}}]{Shim2011}%
  \BibitemOpen
  \bibfield  {author} {\bibinfo {author} {\bibfnamefont {J.-H.}\ \bibnamefont {Shim}}, \bibinfo {author} {\bibfnamefont {H.-G.}\ \bibnamefont {Piao}}, \bibinfo {author} {\bibfnamefont {S.~H.}\ \bibnamefont {Lee}}, \bibinfo {author} {\bibfnamefont {S.~K.}\ \bibnamefont {Oh}}, \bibinfo {author} {\bibfnamefont {S.-C.}\ \bibnamefont {Yu}}, \bibinfo {author} {\bibfnamefont {S.~K.}\ \bibnamefont {Han}},\ and\ \bibinfo {author} {\bibfnamefont {D.-H.}\ \bibnamefont {Kim}},\ }\bibfield  {title} {\bibinfo {title} {Nonlinear motion of magnetic vortex under alternating-current magnetic field: Dynamic correction of a gyrovector and a damping tensor of the thiele's equation},\ }\href {https://doi.org/10.1063/1.3645595} {\bibfield  {journal} {\bibinfo  {journal} {Appl. Phys. Lett.}\ }\textbf {\bibinfo {volume} {99}},\ \bibinfo {pages} {142505} (\bibinfo {year} {2011})}\BibitemShut {NoStop}%
\bibitem [{\citenamefont {V\"olkel}\ \emph {et~al.}(1994)\citenamefont {V\"olkel}, \citenamefont {Wysin}, \citenamefont {Mertens}, \citenamefont {Bishop},\ and\ \citenamefont {Schnitzer}}]{Volkel1994}%
  \BibitemOpen
  \bibfield  {author} {\bibinfo {author} {\bibfnamefont {A.~R.}\ \bibnamefont {V\"olkel}}, \bibinfo {author} {\bibfnamefont {G.~M.}\ \bibnamefont {Wysin}}, \bibinfo {author} {\bibfnamefont {F.~G.}\ \bibnamefont {Mertens}}, \bibinfo {author} {\bibfnamefont {A.~R.}\ \bibnamefont {Bishop}},\ and\ \bibinfo {author} {\bibfnamefont {H.~J.}\ \bibnamefont {Schnitzer}},\ }\bibfield  {title} {\bibinfo {title} {Collective-variable approach to the dynamics of nonlinear magnetic excitations with application to vortices},\ }\href {https://doi.org/10.1103/PhysRevB.50.12711} {\bibfield  {journal} {\bibinfo  {journal} {Phys. Rev. B}\ }\textbf {\bibinfo {volume} {50}},\ \bibinfo {pages} {12711} (\bibinfo {year} {1994})}\BibitemShut {NoStop}%
\bibitem [{\citenamefont {Wysin}(1996)}]{Wysin1996}%
  \BibitemOpen
  \bibfield  {author} {\bibinfo {author} {\bibfnamefont {G.~M.}\ \bibnamefont {Wysin}},\ }\bibfield  {title} {\bibinfo {title} {Magnetic vortex mass in two-dimensional easy-plane magnets},\ }\href {https://doi.org/10.1103/PhysRevB.54.15156} {\bibfield  {journal} {\bibinfo  {journal} {Phys. Rev. B}\ }\textbf {\bibinfo {volume} {54}},\ \bibinfo {pages} {15156} (\bibinfo {year} {1996})}\BibitemShut {NoStop}%
\bibitem [{\citenamefont {Chen}\ \emph {et~al.}(2022)\citenamefont {Chen}, \citenamefont {Abreu~Araujo}, \citenamefont {Riou}, \citenamefont {Torrejon}, \citenamefont {Ravelosona}, \citenamefont {Kang}, \citenamefont {Zhao}, \citenamefont {Grollier},\ and\ \citenamefont {Querlioz}}]{Chen2022}%
  \BibitemOpen
  \bibfield  {author} {\bibinfo {author} {\bibfnamefont {X.}~\bibnamefont {Chen}}, \bibinfo {author} {\bibfnamefont {F.}~\bibnamefont {Abreu~Araujo}}, \bibinfo {author} {\bibfnamefont {M.}~\bibnamefont {Riou}}, \bibinfo {author} {\bibfnamefont {J.}~\bibnamefont {Torrejon}}, \bibinfo {author} {\bibfnamefont {D.}~\bibnamefont {Ravelosona}}, \bibinfo {author} {\bibfnamefont {W.}~\bibnamefont {Kang}}, \bibinfo {author} {\bibfnamefont {W.}~\bibnamefont {Zhao}}, \bibinfo {author} {\bibfnamefont {J.}~\bibnamefont {Grollier}},\ and\ \bibinfo {author} {\bibfnamefont {D.}~\bibnamefont {Querlioz}},\ }\bibfield  {title} {\bibinfo {title} {Forecasting the outcome of spintronic experiments with neural ordinary differential equations},\ }\href {https://doi.org/10.1038/s41467-022-28571-7} {\bibfield  {journal} {\bibinfo  {journal} {Nat. Commun.}\ }\textbf {\bibinfo {volume} {13}},\ \bibinfo {pages} {1016} (\bibinfo {year} {2022})}\BibitemShut {NoStop}%
\bibitem [{\citenamefont {Moreels}\ \emph {et~al.}(2026)\citenamefont {Moreels}, \citenamefont {Lateur}, \citenamefont {De~Gusem}, \citenamefont {Mulkers}, \citenamefont {Maes}, \citenamefont {Milo{\v{s}}evi{\'{c}}}, \citenamefont {Leliaert},\ and\ \citenamefont {Van~Waeyenberge}}]{Moreels2026}%
  \BibitemOpen
  \bibfield  {author} {\bibinfo {author} {\bibfnamefont {L.}~\bibnamefont {Moreels}}, \bibinfo {author} {\bibfnamefont {I.}~\bibnamefont {Lateur}}, \bibinfo {author} {\bibfnamefont {D.}~\bibnamefont {De~Gusem}}, \bibinfo {author} {\bibfnamefont {J.}~\bibnamefont {Mulkers}}, \bibinfo {author} {\bibfnamefont {J.}~\bibnamefont {Maes}}, \bibinfo {author} {\bibfnamefont {M.~V.}\ \bibnamefont {Milo{\v{s}}evi{\'{c}}}}, \bibinfo {author} {\bibfnamefont {J.}~\bibnamefont {Leliaert}},\ and\ \bibinfo {author} {\bibfnamefont {B.}~\bibnamefont {Van~Waeyenberge}},\ }\bibfield  {title} {\bibinfo {title} {mumax+: extensible gpu-accelerated micromagnetics and beyond},\ }\href {https://doi.org/10.1038/s41524-025-01893-y} {\bibfield  {journal} {\bibinfo  {journal} {npj Comput. Mater.}\ }\textbf {\bibinfo {volume} {12}},\ \bibinfo {pages} {71} (\bibinfo {year} {2026})}\BibitemShut {NoStop}%
\bibitem [{\citenamefont {Abreu~Araujo}\ \emph {et~al.}(2022)\citenamefont {Abreu~Araujo}, \citenamefont {Chopin},\ and\ \citenamefont {de~Wergifosse}}]{AbreuAraujo2022}%
  \BibitemOpen
  \bibfield  {author} {\bibinfo {author} {\bibfnamefont {F.}~\bibnamefont {Abreu~Araujo}}, \bibinfo {author} {\bibfnamefont {C.}~\bibnamefont {Chopin}},\ and\ \bibinfo {author} {\bibfnamefont {S.}~\bibnamefont {de~Wergifosse}},\ }\bibfield  {title} {\bibinfo {title} {Amp{\`e}re-oersted field splitting of the nonlinear spin-torque vortex oscillator dynamics},\ }\href {https://doi.org/10.1038/s41598-022-14574-3} {\bibfield  {journal} {\bibinfo  {journal} {Sci. Rep.}\ }\textbf {\bibinfo {volume} {12}},\ \bibinfo {pages} {10605} (\bibinfo {year} {2022})}\BibitemShut {NoStop}%
\bibitem [{\citenamefont {Thiaville}\ and\ \citenamefont {Nakatani}(2006)}]{Thiaville2006}%
  \BibitemOpen
  \bibfield  {author} {\bibinfo {author} {\bibfnamefont {A.}~\bibnamefont {Thiaville}}\ and\ \bibinfo {author} {\bibfnamefont {Y.}~\bibnamefont {Nakatani}},\ }\bibfield  {title} {\bibinfo {title} {Domain-wall dynamics in nanowires and nanostrips},\ }in\ \href {https://doi.org/10.1007/10938171_5} {\emph {\bibinfo {booktitle} {Spin Dynamics in Confined Magnetic Structures III}}},\ \bibinfo {editor} {edited by\ \bibinfo {editor} {\bibfnamefont {B.}~\bibnamefont {Hillebrands}}\ and\ \bibinfo {editor} {\bibfnamefont {A.}~\bibnamefont {Thiaville}}}\ (\bibinfo  {publisher} {Springer Berlin Heidelberg},\ \bibinfo {year} {2006})\ pp.\ \bibinfo {pages} {161--205}\BibitemShut {NoStop}%
\bibitem [{\citenamefont {Makhfudz}\ \emph {et~al.}(2012)\citenamefont {Makhfudz}, \citenamefont {Kr{\"u}ger},\ and\ \citenamefont {Tchernyshyov}}]{Makhfudz2012}%
  \BibitemOpen
  \bibfield  {author} {\bibinfo {author} {\bibfnamefont {I.}~\bibnamefont {Makhfudz}}, \bibinfo {author} {\bibfnamefont {B.}~\bibnamefont {Kr{\"u}ger}},\ and\ \bibinfo {author} {\bibfnamefont {O.}~\bibnamefont {Tchernyshyov}},\ }\bibfield  {title} {\bibinfo {title} {Inertia and chiral edge modes of a skyrmion magnetic bubble},\ }\href {https://doi.org/10.1103/PhysRevLett.109.217201} {\bibfield  {journal} {\bibinfo  {journal} {Phys. Rev. Lett.}\ }\textbf {\bibinfo {volume} {109}},\ \bibinfo {pages} {217201} (\bibinfo {year} {2012})}\BibitemShut {NoStop}%
\bibitem [{\citenamefont {Sch{\"u}tte}\ \emph {et~al.}(2014)\citenamefont {Sch{\"u}tte}, \citenamefont {Iwasaki}, \citenamefont {Rosch},\ and\ \citenamefont {Nagaosa}}]{Schuette2014}%
  \BibitemOpen
  \bibfield  {author} {\bibinfo {author} {\bibfnamefont {C.}~\bibnamefont {Sch{\"u}tte}}, \bibinfo {author} {\bibfnamefont {J.}~\bibnamefont {Iwasaki}}, \bibinfo {author} {\bibfnamefont {A.}~\bibnamefont {Rosch}},\ and\ \bibinfo {author} {\bibfnamefont {N.}~\bibnamefont {Nagaosa}},\ }\bibfield  {title} {\bibinfo {title} {Inertia, diffusion, and dynamics of a driven skyrmion},\ }\href {https://doi.org/10.1103/PhysRevB.90.174434} {\bibfield  {journal} {\bibinfo  {journal} {Phys. Rev. B}\ }\textbf {\bibinfo {volume} {90}},\ \bibinfo {pages} {174434} (\bibinfo {year} {2014})}\BibitemShut {NoStop}%
\bibitem [{\citenamefont {Wang}\ \emph {et~al.}(2019)\citenamefont {Wang}, \citenamefont {Qaiumzadeh},\ and\ \citenamefont {Brataas}}]{Wang2019}%
  \BibitemOpen
  \bibfield  {author} {\bibinfo {author} {\bibfnamefont {X.~S.}\ \bibnamefont {Wang}}, \bibinfo {author} {\bibfnamefont {A.}~\bibnamefont {Qaiumzadeh}},\ and\ \bibinfo {author} {\bibfnamefont {A.}~\bibnamefont {Brataas}},\ }\bibfield  {title} {\bibinfo {title} {Current-driven dynamics of magnetic hopfions},\ }\href {https://doi.org/10.1103/PhysRevLett.123.147203} {\bibfield  {journal} {\bibinfo  {journal} {Phys. Rev. Lett.}\ }\textbf {\bibinfo {volume} {123}},\ \bibinfo {pages} {147203} (\bibinfo {year} {2019})}\BibitemShut {NoStop}%
\bibitem [{\citenamefont {Guslienko}\ \emph {et~al.}(2014)\citenamefont {Guslienko}, \citenamefont {Sukhostavets},\ and\ \citenamefont {Berkov}}]{Guslienko2014}%
  \BibitemOpen
  \bibfield  {author} {\bibinfo {author} {\bibfnamefont {K.~Y.}\ \bibnamefont {Guslienko}}, \bibinfo {author} {\bibfnamefont {O.~V.}\ \bibnamefont {Sukhostavets}},\ and\ \bibinfo {author} {\bibfnamefont {D.~V.}\ \bibnamefont {Berkov}},\ }\bibfield  {title} {\bibinfo {title} {Nonlinear magnetic vortex dynamics in a circular nanodot excited by spin-polarized current},\ }\href {https://doi.org/10.1186/1556-276X-9-386} {\bibfield  {journal} {\bibinfo  {journal} {Nanoscale Res. Lett.}\ }\textbf {\bibinfo {volume} {9}},\ \bibinfo {pages} {386} (\bibinfo {year} {2014})}\BibitemShut {NoStop}%
\bibitem [{\citenamefont {Alfeld}(1984)}]{Alfeld1984}%
  \BibitemOpen
  \bibfield  {author} {\bibinfo {author} {\bibfnamefont {P.}~\bibnamefont {Alfeld}},\ }\bibfield  {title} {\bibinfo {title} {A trivariate clough--tocher scheme for tetrahedral data},\ }\href {https://doi.org/10.1016/0167-8396(84)90029-3} {\bibfield  {journal} {\bibinfo  {journal} {Comput. Aided Geom. Des.}\ }\textbf {\bibinfo {volume} {1}},\ \bibinfo {pages} {169} (\bibinfo {year} {1984})}\BibitemShut {NoStop}%
\bibitem [{\citenamefont {Guslienko}\ \emph {et~al.}(2011)\citenamefont {Guslienko}, \citenamefont {Aranda},\ and\ \citenamefont {Gonzalez}}]{Guslienko2011}%
  \BibitemOpen
  \bibfield  {author} {\bibinfo {author} {\bibfnamefont {K.~Y.}\ \bibnamefont {Guslienko}}, \bibinfo {author} {\bibfnamefont {G.~R.}\ \bibnamefont {Aranda}},\ and\ \bibinfo {author} {\bibfnamefont {J.}~\bibnamefont {Gonzalez}},\ }\bibfield  {title} {\bibinfo {title} {Spin torque and critical currents for magnetic vortex nano-oscillator in nanopillars},\ }\href {https://doi.org/10.1088/1742-6596/292/1/012006} {\bibfield  {journal} {\bibinfo  {journal} {J. Phys.: Conf. Ser.}\ }\textbf {\bibinfo {volume} {292}},\ \bibinfo {pages} {012006} (\bibinfo {year} {2011})}\BibitemShut {NoStop}%
\end{thebibliography}
\end{document}